# MOSMEDDATA: CHEST CT SCANS WITH COVID-19 RELATED FINDINGS DATASET

S.P. Morozov, A.E. Andreychenko, N.A. Pavlov, A.V. Vladzymyrskyy, N.V. Ledikhova, V.A. Gombolevskiy, I.A. Blokhin, P.B. Gelezhe, A.V. Gonchar, V.Yu. Chernina

*Research and Practical Clinical Center of Diagnostics and Telemedicine Technologies, Department of Health Care of Moscow, Russian Federation[1]*

**Abstract**

This dataset contains anonymised human lung computed tomography (CT) scans with COVID-19 related findings, as well as without such findings. A small subset of studies has been annotated with binary pixel masks depicting regions of interests (ground-glass opacifications and consolidations). CT scans were obtained between 1st of March, 2020 and 25th of April, 2020, and provided by municipal hospitals in Moscow, Russia. Permanent link: https://mosmed.ai/datasets/covid19_1110. This dataset is licensed under a Creative Commons Attribution-NonCommercial-NoDerivs 3.0 Unported (CC BY-NC-ND 3.0) License.

**Key words:** artificial intelligence, COVID-19, machine learning, dataset, CT, chest, imaging

**Background**

During the COVID-19 pandemic, most countries faced a tremendous increase in the healthcare burden. This situation calls for the most careful use of financial and human resources than ever before. Unfortunately, the preventive measures put in place in medical facilities are not always enough to avoid deaths of medical workers. The loss of trained specialists in emergency care, radiology, etc., is especially concerning. Computed tomography is considered a key tool to diagnose and evaluate the progression of COVID-19. The CT studies are performed in the outpatient setting and target patients with acute respiratory symptoms, and those with established diagnosis and mild disease progression who are able to recover in their homes (under supervision via telemedicine). Inpatient facilities use CT for initial and differential diagnostic assessment, evaluation of disease progression, and determining, whether the patient should be admitted to the intensive care unit or discharged [1,3,4].

The ever-increasing use of CT translates to an immense burden on the health system. For example, in Moscow, the chain of municipal outpatient CT centers perform about 90 studies per 1 CT scanner per day (the record holder scanner performed 163 studies during one day).

To standardize and streamline clinical decision-making the experts developed a classification model that grades the severity of lung tissue abnormalities observed on CT images along with other symptoms (see Table).

Increased burnout and high risk of occupational death among healthcare workers call for automation of the reading process, which will improve productivity and minimize errors [8]. Preliminary figures indicate the AI algorithms have sufficient accuracy for the diagnostic evaluation of COVID-19 (responsiveness – 90%, specificity – 96%, AUC – 0.96, overall accuracy 76.37-98.26) [6,9].

**Dataset**

Data were obtained between 1st of March, 2020 and 25th of April, 2020, and provided by municipal hospitals in Moscow, Russia. This dataset contains anonymised human lung computed tomography (CT) scans with COVID-19 related findings (CT1-CT4), as well as without such findings (CT0) (fig.).

---

[1] Correspondence: morozov@npcmr.ru. Research and Practical Clinical Center of Diagnostics and Telemedicine Technologies, Department of Health Care of Moscow, 28-1, ul. Srednyaya Kalitnikovskaya, Moscow 109029, Russia. http://www.mosmed.ai.

There are 1110 studies in dataset. Population: 1110 persons, males – 42%, females – 56%, other/unknown – 2%; age from 18 to 97 years, median – 47 years. As a first step, all studies (n=1110) were distributed into 5 categories according to classification (table). Number of cases by category: CT-0 – 254 (22,8%), CT-1 – 684 (61,6%), CT-2 – 125 (11,3%), CT-3 – 45 (4,1%), CT-4 – 2 (0,2%). Secondly, every study has been saved in NifTI format and archived into Gzip archive. During this process only every 10-th image (Instance) was preserved in the final study file.

*Table. Classification of the severity of lung tissue abnormalities with COVID-19 and routing rules*

| Severity | CT | Clinical Data | Decision |
|---|---|---|---|
| Zero | **CT-0** Not consistent with pneumonia (including COVID-19). | – | Inform the attending physician. Refer to a specialist. |
| Mild | **CT-1** Ground glass opacities. Pulmonary parenchymal involvement =<25% OR absence CT signs in typical clinical manifestations and relevant epidemiological history. | A. $t^0 < 38.0\,^0C$<br>B. RR <20/min<br>C. $SpO_2$ >95% | Follow-up at home using telemedicine technologies (mandatory telemonitoring) |
| Moderate | **CT-2** Ground glass opacities. Pulmonary parenchymal involvement 25-50%. | A. $t^0 > 38.5\,^0C$<br>B. RR 20-30/min<br>C. $SpO_2$ 95% | Follow-up at home by a primary care physician |
| Severe | **CT-3** Ground glass opacities. Pulmonary consolidation. Pulmonary parenchymal involvement 50-75%. Lung involvement increased in 24–48 hours by 50% in respiratory impairment per follow-up studies. | One or more signs on the background of fever:<br>A. $t^0 > 38,5\,^0C$<br>B. RR ≥30/min<br>C. $SpO_2$ ≤95%<br>D. Partial pressure of oxygen ($PaO_2$)/ Fraction of inspired oxygen ($FiO_2$) ≤300 mmHg<br>(1 mmHg=0,133 kPa) | Immediate admission to a COVID-specialized hospital. In a hospital setting: immediate transfer to the intensive care and resuscitation unit. Emergency computed tomography (if not done before). |
| Critical | **CT-4** Diffuse ground glass opacities with consolidations and reticular changes. Hydrothorax (bilateral, more on the left). Pulmonary parenchymal involvement >=75%. | Signs of shock, multiple organ failure, respiratory failure. | Emergency medical care. Immediate admission to a specialized hospital for patients diagnosed with COVID-19. In a hospital setting: immediate transfer to the intensive care and resuscitation unit. Emergency computed tomography (if not done before and when patient status allows for it). |

A small subset of studies (n=50) has been annotated by the experts of Research and Practical Clinical Center for Diagnostics and Telemedicine Technologies of the Moscow Health Care Department. During the annotation for every given image ground-glass opacifications and regions of consolidation were selected as positive (white) pixels on the corresponding binary pixel mask. The resulting masks have been saved in NIfTI format and then transformed into Gzip archives. The MedSeg® web-based annotation software has been used for creating binary masks (© 2020 Artificial Intelligence AS).

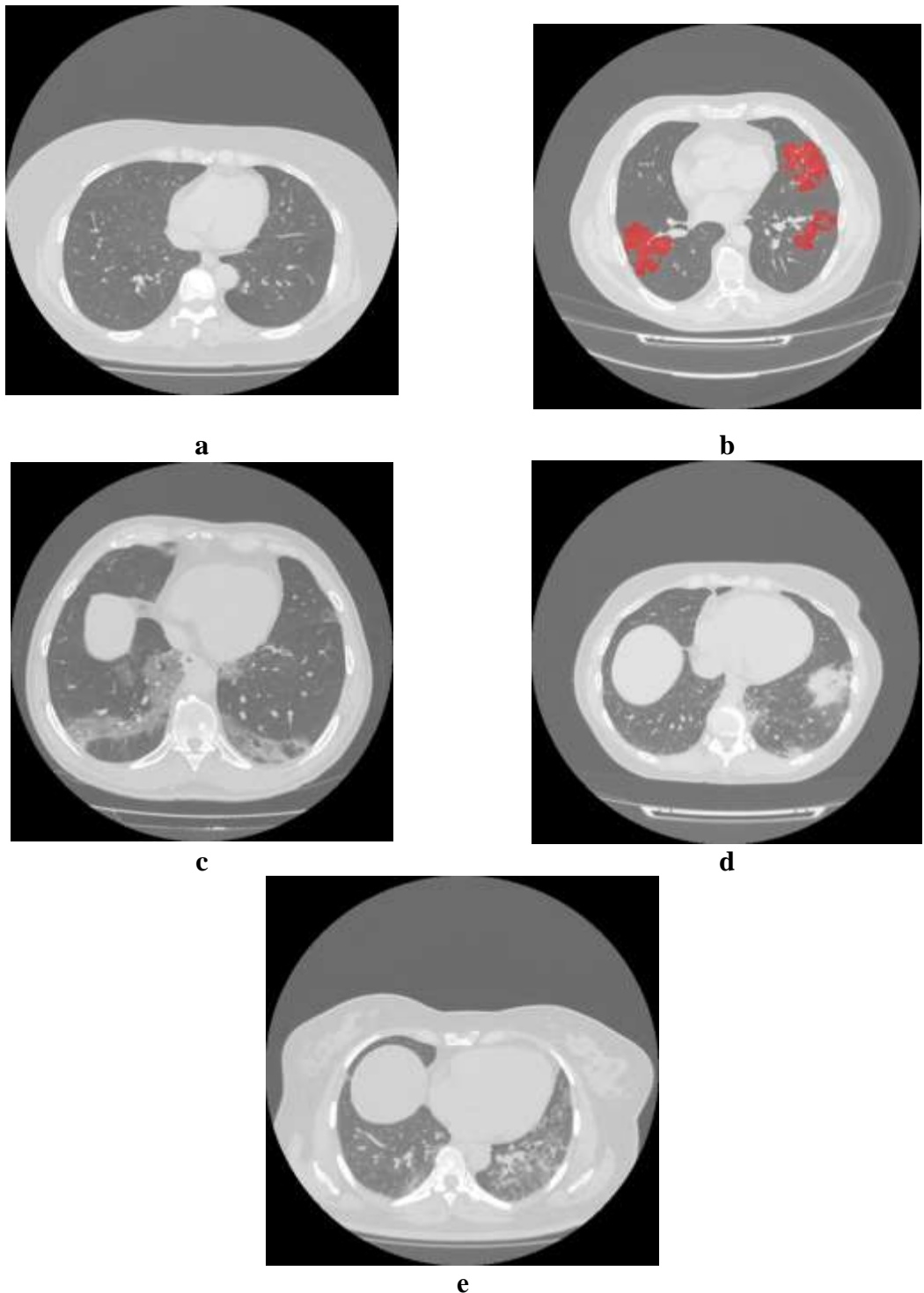

*Fig. Examples of images: a – CT-0, b – CT-1 (overlay: binary mask), c – CT-2, d – CT-3, e – CT-4*

**Value**

The dataset is intended for education, calibration, and independent assessment of the AI (computer vision) algorithms [7].

To help combating COVID-19 the AI (computer vision) algorithms will allow:

1. Triaging patients in outpatient facilities to secure rapid and consistent routing (incl. based on the CT0-4 criteria)

2. Prioritizing studies that contain signs of COVID-19 in the worklist

3. Performing a rapid and high-quality assessment of abnormal changes by comparing several studies

4. Minimizing the risks of errors and missed abnormalities.

At the moment, there is a wide range of publicly available COVID-19 datasets [2,5]. However, this should not be viewed as an obstacle for the following reasons:

1. Development of algorithms requires large volumes of high-quality clinical data that is representative of real-world patient populations

2. The algorithms must be verified with new data, i.e. new datasets that were not used during the learning and calibration phases. The more data are available in the open sources, the better for the developers

3. The available datasets are relatively small and rarely present additional information, such as tags and/or binary masks for the regions of interest (ROI).

Permanent link: https://mosmed.ai/datasets/covid19_1110. This dataset is licensed under a Creative Commons Attribution-NonCommercial-NoDerivs 3.0 Unported (CC BY-NC-ND 3.0) License.